\documentclass[aps,twocolumn,superscriptaddress,showpacs,floatfix]{revtex4-1}
\usepackage{amsfonts,amssymb,latexsym,xspace,epsfig,graphicx,color}
\usepackage{amsmath,enumerate,stmaryrd,xy,stackrel,ulem}

\usepackage{xcolor}

\begin{document}
\title{Current production in ring condensates with a weak link}
\author{Axel P\'erez-Obiol}
\affiliation{Laboratory of Physics, Kochi University of Technology, Tosa Yamada, Kochi 782-8502, Japan}
\author{Juan Polo}
\affiliation{Quantum Systems Unit, Okinawa Institute of Science and Technology Graduate University, Onna, Okinawa 904-0495, Japan}
\author{Taksu Cheon}
\affiliation{Laboratory of Physics, Kochi University of Technology, Tosa Yamada, Kochi 782-8502, Japan}
\date{\today}

\begin{abstract}

We consider attractive and repulsive condensates in a ring trap stirred by a weak link, and analyze the spectrum of solitonic trains dragged by the link, by means of analytical expressions for the wave functions, energies and currents. The precise evolution of current production and destruction in terms of defect formation in the ring and in terms of stirring is studied. We find that any excited state can be coupled to the ground state through two proposed methods: either by adiabatically tuning the link's strength and velocity through precise cycles which avoid the critical velocities and thus unstable regions, or by keeping the link still while setting an auxiliary potential and imprinting a nonlinear phase as the potential is turned off. We also analyze hysteresis cycles through the spectrum of energies and currents.
\end{abstract}

\maketitle

\section{Introduction}

Condensates in ring geometries present a
wealth of superfluid and nonlinear effects 
and yield the potential for the development of atomtronic
devices \cite{Amico_2017}.
Production and decay of supercurrents,
supersonic flow,
hysteresis cycles, and the ability
to sustain solitonic solutions have been widely studied
theoretically and
experimentally~\cite{ryu07,moulder12,beattie13,wright13pra,eckel14,ryu13,hallwood10,schenke11,amico14,Hou_2017,perrin20}.
The production and control of supercurrents is a crucial step towards future quantum devices.
For instance, the atomic analog of the superconducting quantum interference device,
which was realized experimentally in \cite{Wright_2013}, is based on the stirring of
a weak link across a condensate trapped in a ring geometry. This process has been
extensively investigated  and it has been shown capable to produce superposition
of current states \cite{Amico_2015,Aghamalyan_2015}.

One dimensional rings offer the opportunity to analyze
the spectrum much more precisely, and to tackle new effects
which are in general masked by higher dimensional dynamics,
such as non-vortex-antivortex phase slips \cite{piazza09}.
Experimentally, 
the production of currents can be induced by
a rotating weak link. The link produces a low density region and can  be rotated to stir the condensate and produce
superfluid currents~\cite{ramanathan11,piazza09,piazza13,wright13prl}.
Alternatively, a phase can be directly imprinted
on the condensate~\cite{dalibard11,perrin18}. In the latter case, however,
Bose-Josephson Junction (BJJ) oscillations are found in the case of a large enough defect
or a small enough nonlinearity~\cite{polo19}.

With the aim to better understand the behavior of currents
in ring condensates, various works have analyzed these systems
in the mean field limit and at zero temperature.
Current dynamics  have been 
studied through either a rotational
drive~\cite{kanamoto09}, through the interaction between
symmetry breaking potentials and rotation such as in 
lattice rings~\cite{fialko12,li12,munoz19},
or through rotating defects~\cite{baharian13,munoz15,kunimi18}.
Solutions of the Gross-Pitevskii equation (GPE) for a 1D ring,
in the free case
and with various sets of potentials, have been established
by analyzing its spectrum either numerically
and/or through the use of Jacobi elliptic 
functions~\cite{carr002,seaman05,hakim97,pavloff02,cominotti14,shamriz18,perezobiol19}.
The spectrum for a moving link and repulsive interactions was analyzed
thoroughly in~\cite{perezobiol20}. Studying how current states
are coupled to either the ground state or dark solitonic
states, which are found to trigger phase slips,
has proven essential to understand production and decay
of currents, and how to build more robust states.

In this article we complement previous studies by
determining and describing the spectrum and critical velocities 
for both attractive and repulsive stirred condensates.
The use of analytical solutions releases us from the limitation
to study the ground state, and also allows us to explore 
the current dynamics of stirred excited states.
We focus on three main mechanisms for current production:
adiabatic excitation, hysteresis, and phase imprinting. 
In the case of adiabatic excitation, we distinguish two types of stirrings,
one which starts at zero velocity, and another in which the link is
set while rotating, allowing for production of larger currents.
Each mechanism is thoroughly analyzed through the spectrum of energies
and currents. We provide explicit protocols to produce the first
excited states, and  compare the cases for repulsive and attractive interactions; to the best of our knowledge, this is the first time this analysis has been performed for attractive interactions.

This article is organized as follows. The main features of the spectrum 
are laid out in Sec.~\ref{sec:solutions},
and details are given in Appendix~\ref{app:spectrum}.
In Sec.~\ref{sec:current}, we analyze how currents depend on the link's
velocity and strength, through regular
stirring, in Sec. \ref{sec:regular},
and through a set of adiabatic cycles which
are able to couple any stationary current to the ground state, in Sec. \ref{sec:excitation}.
We also connect the energy and current spectra to a set of hysteresis
cycles in Sec.~\ref{sec:hysteresis}.
In Sec.~\ref{sec:auxiliary}, we present an alternative method for stable
current production in rings with weak links.
This protocol does not involve the movement of the link, but 
setting an auxiliary potential and phase imprinting
a nonlinear slope, so that no BJJ oscillations are found.
We conclude in Sec.~\ref{sec:conclusions}.

\section{Exact spectrum of a stirred BEC}
\label{sec:solutions}

\begin{figure}[t]
\centering
\includegraphics[width=.48\textwidth,height=.6\textwidth]{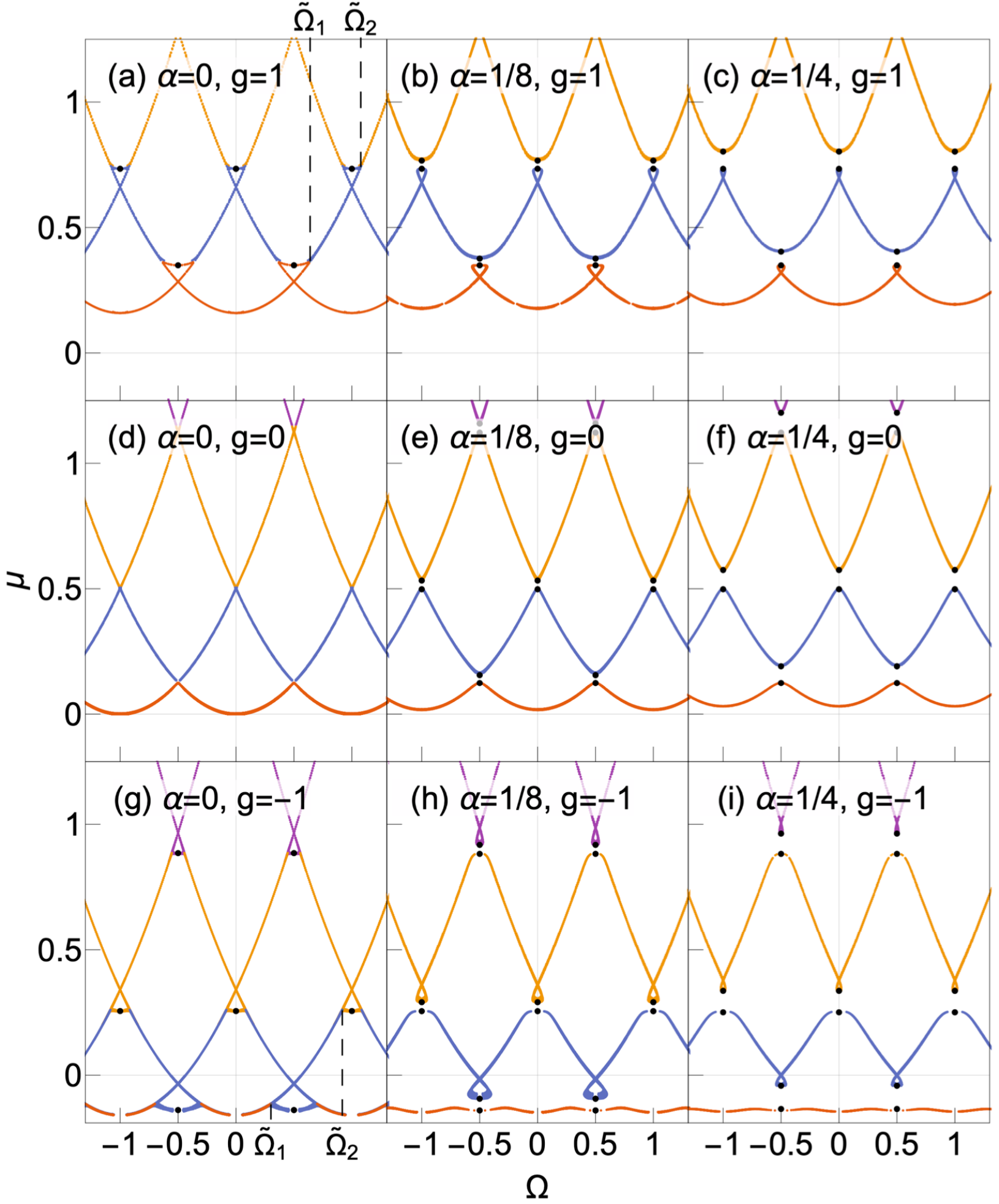}
\caption{Spectrum of energies $\mu$ of solitonic solutions moving at
constant velocity $\Omega$, in the frame of reference
of the moving waves and/or link.
Left-hand side plots correspond to free solutions on a ring. The parabolas centered at 
$\Omega=0$
in panels (a), (d), and (g) correspond to the energy of the ground state
as observed in the frame of reference moving at $\Omega$.
The other parabolas are the energies of vortex states in the same frame of reference.
The flat lines crossing among parabolas in each of these plots correspond to gray solitonic trains, where the bottom ones correspond to densities with a single dip, and each
upper line to a solitonic train with one more dip in the density.
Middle and right-hand side columns correspond to energies of solitonic trains dragged
by a link of strengths $\alpha=\frac18,\frac14$. 
Each set of concatenated swallowtails
is plotted in a different color.
Red, blue, and orange bands (first, second and third from the bottom)
 correspond to solitonic trains with one, two, and three dips in the density.
The plots in the middle column continuously turn into the left column
ones as $\alpha$ decreases to zero. The colors in the left panels and 
the velocities $\tilde{\Omega}_1$, $\tilde{\Omega}_2$
indicate how solutions are split into separate bands if a weak link
is turned on, i.e. how they are coupled to the middle and right plots through
a variation of $\alpha$.
Solid black dots correspond to dark solitonic trains, where the density
minima are zero. Units are dimensionless, with $R=M=\hbar=1$.
}
\label{fig:st3x3}
\end{figure}

Within the mean field limit, and at zero temperature,
the condensate wave function on a ring $\psi(\theta,t)$ is determined
by the Gross-Pitaevskii equation (GPE),
\begin{align}
 i\hbar\,\partial_t \psi(\theta,t)=&
-\frac{\hbar^2}{2M R^2}\partial_\theta^2\psi(\theta,t)
+g|\psi(\theta,t)|^2\psi(\theta,t)
\nonumber\\\
&+V(\theta,t)\psi(\theta,t),
\end{align}
with $\theta\in[0,2\pi)$, $g$ the reduced 1D coupling,
and $V(\theta)$ an external potential.
From here onwards we work in natural units, where 
the ring's radius $R$, the atomic mass $M$,
and $\hbar$ are $R=M=\hbar=1$, and
in the frame of reference comoving with the link.
In this frame of reference, where the link
is modeled by a static Dirac delta,
the stationary wave function $\phi(\theta)$ and chemical potential $\mu$
are fully determined by
\begin{align}
\label{eq:gpf}
-\frac12 \phi''(\theta)+g|\phi(\theta)|^2\phi(\theta)=&\mu\,\phi(\theta),
\\
\label{eq:bc1}
\phi(0)-e^{i 2\pi\Omega}\phi(2\pi)=&0,
\\
\label{eq:bc2}
\phi'(0)-e^{i2\pi\Omega}\phi'(2\pi)=&2\alpha\,\phi(0),
\end{align}
where $\alpha$ and $\Omega$ are the link's strength and velocity,
and where the wave function is normalized to
$\int_0^{2\pi}d\theta|\phi(\theta)|^2=1$.
This framework allows us to use analytical expressions for the wave functions,
chemical potentials, and currents.
It takes advantage of the elliptic functions, which appear as solutions
of the nonlinear Schr\"odinger equation. These functions are in general useful in
1D GPE models with point interactions, see for example~\cite{seaman05,kunimi19},
or in developing solvable models which either generalize the GPE
to a higher order nonlinear Schr\"odinger equation~\cite{takahashi16},
or apply Darboux transformations to find solvable potentials~\cite{feng19}.
In our case, the solutions are those of the free GPE
and the moving potential is relegated to applying in them boundary conditions~(\ref{eq:bc1})
and~(\ref{eq:bc2}).
In particular, the chemical potential is given by 
\begin{align}
\label{eq:mu}
\mu(k,m)=&\frac{1}{4\pi}\left(3g+2k^2(m-2)+3k\,\eta\right),
\end{align}
with $k\geq0$ and $m\in[0,1]$ the frequency and elliptic modulus
of the Jacobi solution, and
$\eta=E[{\rm JA}(k(2\pi-\theta_0),m),m]+E[{\rm JA}(k\,\theta_0,m),m]$.
$\theta_0$ is a shift depending on $k$ and $m$ such that the wave is continuous
at $\theta=0,2\pi$,
$E$ is the elliptic integral of the second kind,
and ${\rm JA}$ the Jacobi amplitude.
Equations~(\ref{eq:bc1})-(\ref{eq:mu}) then provide $\alpha(k,m)$,
$\Omega(k,m)$, and $\mu(k,m)$, from which we compute
 $\mu(\alpha,\Omega)$.
The precise relationship between the frequency and the elliptic modulus, $k$ and $m$,
and the weak link strength and velocity, $\alpha$ and $\Omega$, is somewhat
convoluted and not shown here.
For more details we refer to Appendix~\ref{app:spectrum} and~\cite{perezobiol20}.

The full energy spectrum of dragged solitons in the link's frame of reference is plotted in
Fig.~\ref{fig:st3x3} for reduced 1D couplings $g=-1,0,1$ and delta strengths
$\alpha=0,\frac18,\frac14$. The general solution has the form of a
moving gray solitonic train, which may become a dark solitonic train
at $\Omega=n,\frac{n}{2}$, with $n$ integer 
(black dots in Fig.~\ref{fig:st3x3}), or plane waves in the absence of a link, at $\alpha=0$
(parabolas in Fig.~\ref{fig:st3x3} (a), (d) and (g)).
This spectrum is characterized, for either attractive or repulsive condensates,
by a set of concatenated swallowtail diagrams, each forming an energy band
corresponding to  solutions with a different number of
solitons, indicated by different colors (shades of gray) in Fig~\ref{fig:st3x3}.
These energy levels only cross for $\alpha=0$. In this case, the parabolas correspond to plane waves,
and represent the energies of vortex states from the point of view
of an observer moving at $\Omega$. The lines crossing among these parabolas
correspond to gray solitons freely moving at $\Omega$.

The spectrum of dragged solutions for attractive and repulsive condensates
differ qualitatively in two main ways.
Firstly, for $g<0$, swallowtails point upward,
while for $g>0$ they point downward.
This implies that a given stirring protocol produces solutions with a different number of solitons
in repulsive and attractive condensates.
Secondly, the band formed by the ground states at different $\Omega$ for $g<0$
and $\alpha>0$, plotted as the red (bottom, light gray) line in Fig.~\ref{fig:st3x3} (h) and (i),
contains no swallowtail structure and forms a continuous set of solutions.
 Since these solutions correspond to the ground
state of the condensate for each velocity, they are also stable against Bogoliubov
perturbations. This means that, in attractive condensates,
 solutions with one gray soliton, i.e. the dip created by the link,
 and with largely different currents can be coupled
 among them through a simple adiabatic variation of the velocity of the link, $\Omega$.

Each set of concatenated swallowtails contains top parts and bottom parts.
Each of these parts defines a set of solutions
which are continuously connected through a variation of $\Omega$,
and is limited by a pair of velocities which mark the tips of the tails
 (loops) in each diagram, and which we call critical velocities.
 Note that these critical velocities defined here are not the same as the critical velocity
of the fluid which is given by its local speed of sound.  
Beyond these velocities, stationary solutions for the corresponding band
do not exist,
and the condensate is not able to sustain solitons comoving with the link.
Moreover, these velocities mark the point at which top part and bottom part
solutions merge, indicating also instability with respect to Bogoliubov perturbations~\cite{mueller02}.
This was explicitly found for repulsive condensates, where the top parts were found to be unstable
through a Bogoliubov analysis, implying that the stable solutions
from the bottom part also become unstable at the critical velocity,
where they merge with the unstable ones~\cite{perezobiol20}.
In the current production mechanisms discussed in the following,
the Bogoliubov analysis is not as relevant for attractive condensates.
In this case, hysteresis cycles do not exist, 
and adiabatic excitation is possible through paths
that include only ground states of the corresponding Hamiltonians, which are stable by definition,
see Secs.~\ref{sec:excitation} and~\ref{sec:hysteresis}.

The critical velocities are computed through Eqs.~(\ref{eq:gpf})-(\ref{eq:bc2}).
In the limit $\alpha\to0$ and for $g\neq 0$,
where plane wave and gray solitonic solutions merge,
these velocities have a simple analytical form  given by
\begin{equation}
\Omega_c = \pm\tilde{\Omega}_n\pm l,
\end{equation}
with $\tilde{\Omega}_n=\sqrt{\frac{g}{2\pi}+\frac{n^2}{4}}$ and with
two integers $n>0$ and $l\geq0$~\cite{perezobiol20}.
As the barrier strength increases, the critical velocities $\Omega_c$
monotonically decrease. 
In the limit $\alpha\to\infty$, where gray solitons become fully formed dark solitons
with their corresponding phase jump and zero valued density dip, $\Omega_c$ converge
to a value of $\pm\frac{n}{2}\pm l$.

The spectrum plotted in Fig.~\ref{fig:st3x3} is then essential
to qualitatively understand how to avoid critical velocities 
and particular states such as dark solitonic trains
when stirring the condensate.
This is useful for a better control of the condensate
in processes with stirring or moving defects,
such as the proposed atomic superconducting interference device~\cite{Amico_2015}.
It also shows that if an impurity or a link is set at half integer velocity,
 pairs of dark solitonic solutions separated by a narrow gap appear in the spectrum.
Some of these dark solitonic states are found unstable with respect to perturbations
in the condensate through a Bogoliubov analysis~\cite{perezobiol20}, and
are also shown to trigger phase slips~\cite{polo19}.
Finally, Fig.~\ref{fig:st3x3} also illustrates how the different excited
states can be coupled among 
them and to the ground state by tuning
the link strength $\alpha$ and velocity $\Omega$,
providing a basis to study hysteresis cycles.
We illustrate a sample of such stirring protocols in the following sections.

\section{Current production}
\label{sec:current}

The current, $J=-\frac{i}{2}\int d\theta (\phi^*\phi'-\phi{\phi^*}')$, for a link
modeled by a Dirac delta, is given by
\begin{align}
J=\pm 2\pi \gamma+ n,
\end{align}
with $n$ an integer, and
\begin{align}
\gamma=&\frac{1}{g(2\pi)^{3/2}}\sqrt{g+k\,\eta}\sqrt{g-2\pi k^2+k\,\eta}
\nonumber\\&\times\sqrt{g-2\pi k^2(1-m)+k\,\eta}.
\label{eq:gamma}
\end{align}
Together with $\alpha(k,m)$ and $\Omega(k,m)$, 
the current can be found in terms of $\alpha$ and $\Omega$
by scanning the well defined parameter space given by
the frequency $k\geq0$ and elliptic modulus $m\in[0,1]$,
as done with the chemical potential.
The analytical results shown in the following plots are corroborated by
simulations of the time-dependent GPE in the lab frame,
where a peaked Gaussian potential explicitly moves around the ring.
Solutions from both methods
are found to overlap for Gaussian amplitude widths $\sigma=2\pi/200$.

\subsection{Adiabatic, regular stirring}
\label{sec:regular}

\begin{figure}[t]
\centering
\includegraphics[width=.45\textwidth,height=0.365\textwidth]{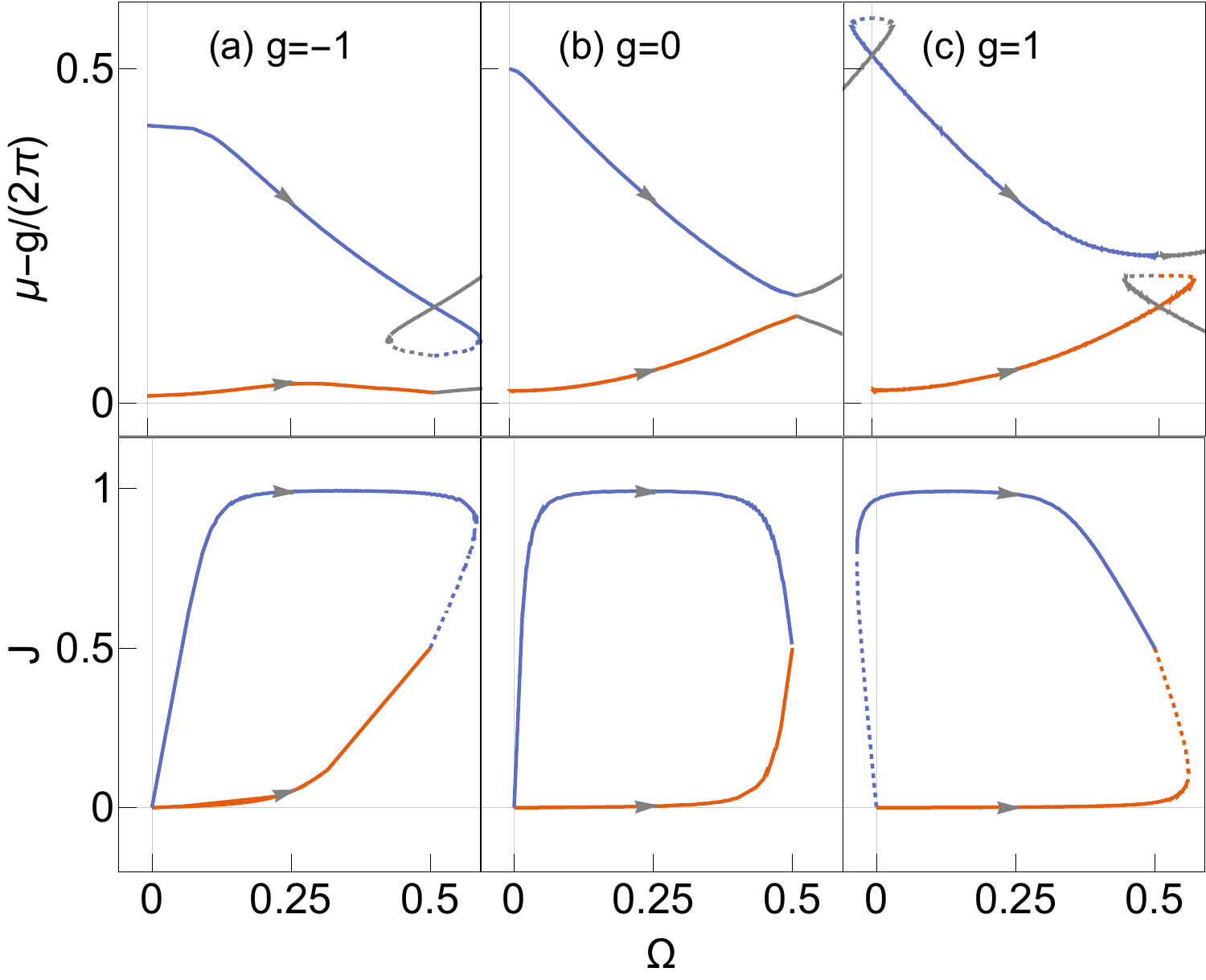}
\caption{Current and energy evolution as the ground state in red (bottom line, light gray), and first
excited state with two gray solitons in blue (top, dark gray), are stirred 
with a link of $\alpha=\frac18$ and velocities
up to $\Omega\simeq 0.5$, and nonlinearities $g=-1,0,1$.
Dashed lines correspond to the bottom
and top parts of upward and downward swallowtail diagrams.
The gray (right) lines in top plots are included to better visualize
the swallowtail structure. All units are in terms of $R=M=\hbar=1$}
\label{fig:1dstirring}
\end{figure}

Figure~\ref{fig:1dstirring} shows the current evolution for three different cases, $g=-1,0,1$,
as a link is set on
the ground (red line/bottom one with lighter shade) and first excited (blue line/second and darker one) state and then stirred
by adiabatically increasing the velocity 
 from $\Omega=0$ to $\Omega\simeq0.5$.

For $g<0$, a link can be set in the ground state
and its velocity increased indefinitely.
The current can be steadily increased,
and the solutions alternate between a shallow gray soliton
moving at $\Omega=n$
and a dark soliton moving at $\Omega=n+\frac12$.
See the red (bottom, light gray) lines in Fig.~\ref{fig:1dstirring} (a) 
for the energy and current evolution in the path $\Omega=0\to 0.5$.
This current is produced more abruptly for attractive interactions closer to zero.
For weak interactions, $g\lesssim0$, the first excited state
consists in two dark solitons, with $J=0$.
When this state is stirred, a current $J\simeq 1$ in the stirring direction
is produced at very small velocities, and then is kept roughly constant
up to the critical velocity, $\Omega_c\gtrsim0.5$, see the blue (top, dark gray) line in Fig.~\ref{fig:1dstirring} (a).
See also Fig.~\ref{fig:densities} for the densities of the
ground and first excited states at $\Omega=0,0.5$.

For the repulsive case, stirring the ground state with increasing velocity
leaves the current practically constant, and a critical velocity is found
at $\Omega\gtrsim 0.5$, $\Omega\in(0.5,\tilde{\Omega}_1)$,
as shown in Fig.~\ref{fig:1dstirring} (c) (red/bottom line).
This velocity marks the tip of the lowest and right swallowtail,
and is connected to
the solution with one dark soliton through
a set of unstable solutions~\cite{perezobiol20}.
The first excited state corresponds to the first vortex
state and for $\alpha\gtrsim0$ contains two gray solitons. 
Its energy is represented by the crossing blue and gray (upper) lines in Fig.~\ref{fig:1dstirring} (c).
In the limit $\alpha\to0$,
it turns into a plane wave with one unit of angular momentum.
If this vortex is stirred, the current remains roughly constant for small velocities,
and rapidly decreases to $J=0.5$ at $\Omega\lesssim 0.5$.

The linear case, $g=0$, in panel (b) of Fig.~\ref{fig:1dstirring},
presents similar current dynamics,
except that initial states are all static ($J=0$), 
and no critical velocities are encountered in any
stirring. In this case, link velocities can be increased and decreased indefinitely.

The adiabatic paths described above can also
be understood in reverse, that is, in terms of a decreasing
stirring velocity.
Moreover, we note that, due to rotational symmetry, the evolution of
the currents of these paths is also valid for the same
states and stirrings but with velocities $\Omega\to\pm\Omega+n$
and currents $J\to \pm J+n$, with $n$ an integer.

\subsection{Adiabatic, excitation stirring}
\label{sec:excitation}

The stirring procedures of Fig.~\ref{fig:1dstirring}, consisting in a steady
increase of the link's velocity up to $\Omega\simeq0.5$,
allow us to produce currents $|\Delta J|\lesssim 1$.
Passed these velocities, critical velocities are encountered,
and the condensate cannot be adiabatically excited anymore.
However, there are cases where critical velocities are not a limitation. In particular,
the ground state of attractive condensates,
any state for the linear case,
and in general dark solitonic states in the limit $\alpha\to\infty$,
where the tails in the energy spectrum shrink and vanish.
These states can be continuously excited to states with larger currents
by constantly increasing the velocity of the link.

\begin{figure}[t]
\centering
\includegraphics[width=.48\textwidth,height=.51\textwidth]{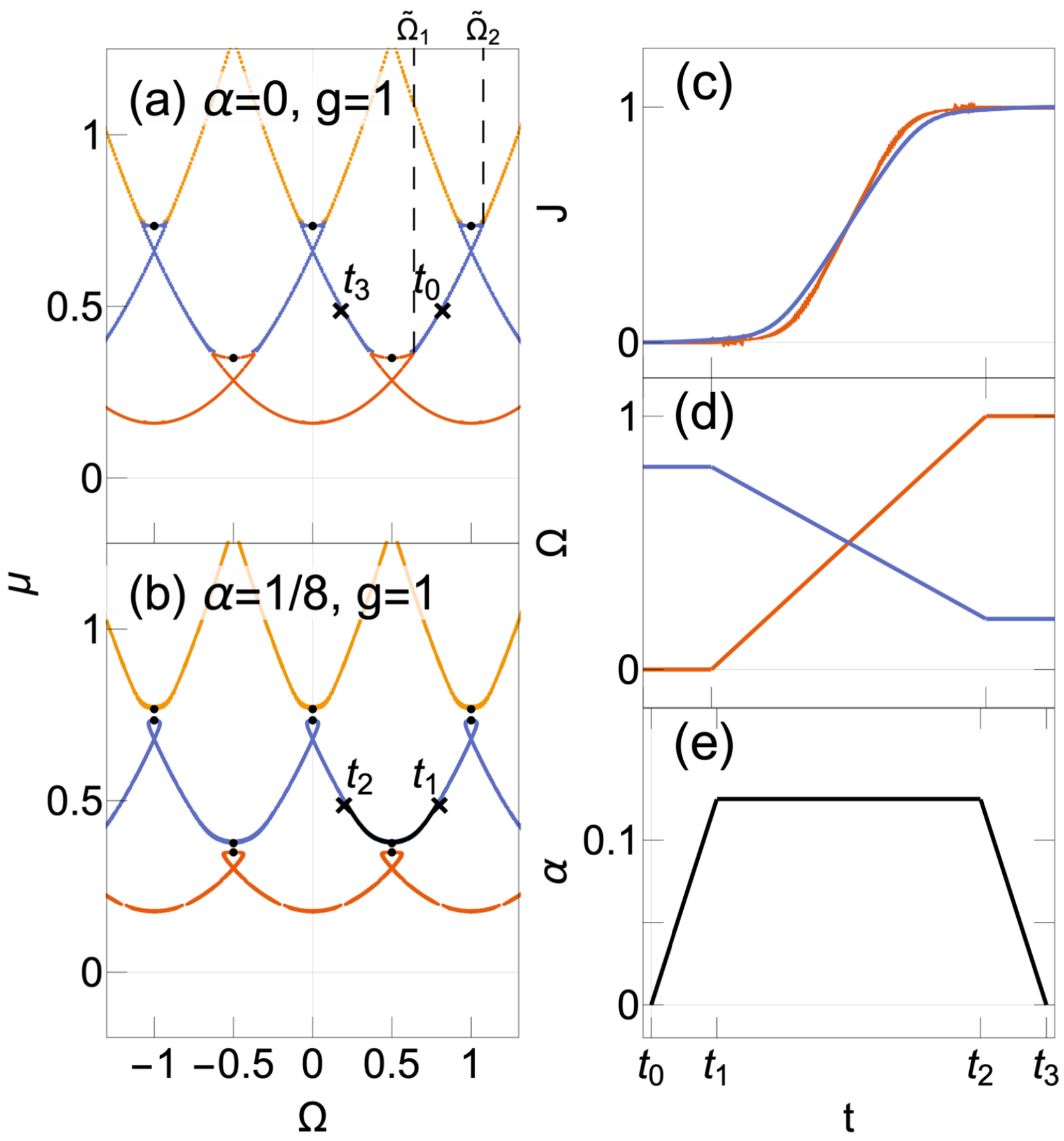}
\caption{
Spectrum of energies for $g=1$ in the absence of a link (a)
and with a link of $\alpha=\frac18$ (b), together with the
key points in the adiabatic path to excite the condensate
to current $J=1$.
Points $t_0$ and $t_3$ correspond to the ground state and first vortex
state in the reference frame of an observer moving at $\Omega_i\in(\tilde{\Omega}_1,\tilde{\Omega}_2)$,
and $\Omega_f=-\Omega_i+1$.
Points $t_1$ and $t_2$, and the path uniting them, consist of solutions
with two gray solitons being dragged by the link.
On the right-hand side plots we show in blue the evolution of the current (c) and
link velocity (d), and in black the link strength (e) between times
$t_0$ and $t_3$. Exciting attractive condensates from the ground state
also involves setting and unsetting a link as in plot (e).
In this case, the link is set at zero velocity and then increased
(red/light-gray line in plot (d)), the current evolving similarly
as for the repulsive condensate (red/light-gray line in plot (c)).
All quantities are in natural units.
}
\label{fig:currentsj1j2}
\end{figure}

We follow similar procedures in attractive
and repulsive condensates
that couple the ground state to excited states so that
$|\Delta J| \geq 1$.
For repulsive condensates, a link is set in the ground state
while rotating at a velocity $\Omega_i$.
Initial velocities $\Omega_i\in(\tilde{\Omega}_n,\tilde{\Omega}_{n+1})$
access the $n^{\rm th}$ excited state, i.e. the one with $n+1$ solitons.
The velocity is then decreased down to the other
side of the swallowtail diagram, and then
the weak link is turned off.
These cycles include an intermediate
dark solitonic solution (with $n+1$ dark solitons), and the current increases more abruptly
in the middle points of the paths.
An example of such cycles that excites the repulsive condensate  to $J=1$
is shown in Fig.~\ref{fig:currentsj1j2}.
At time $t_0$, an observer is rotating at $\Omega_i\in(\tilde{\Omega}_1,\tilde{\Omega}_{2})$
around the ground state. 
As this  observer sets a link while moving at $\Omega$, at point $t_1$,
two dips in the density are created moving at the same velocity,
one at the link's position and another in the opposite site.
As the velocity is decreased,
the dips become deeper, forming a dark solitonic train at $\Omega=0.5$,
and returning to the original gray solitonic train at $t_2$. At this point,
however, the link and gray solitons are moving much slower and 
the condensate current is close to $J=1$.
By removing the slowly moving weak link, we recover the free and flat
condensate, now with a current $J=1$, at point $t_3$.

Attractive states with $g\lesssim0$ allow for two
possible excitation processes. Firstly,
and analogously to the repulsive case,
a link can be set at a finite velocity $\Omega_i$.
Initial rotations
$\Omega_i\in(\tilde{\Omega}_n,\tilde{\Omega}_{n+1})$ 
produce solutions corresponding to the $n^{\rm th}$
excited state.
In this case, the velocity must be {\it increased}
up to the following swallowtail diagram, such that critical velocities
are avoided, see Fig.~\ref{fig:st3x3}.
Secondly, and perhaps more naturally, we can stir by setting
a link at zero velocity, and then speed up the stirring.
This process takes advantage that no critical velocities are encountered
when stirring the ground state of attractive condensates.
Removing the link at $\Omega=n$ will then produce vortex states
with $n$ quanta of angular momentum.
The evolution of the current up to $J=1$,
and of the link's velocity and strength
corresponding to this protocol, are plotted in Fig.~\ref{fig:currentsj1j2}
(c), (d) and (e).

We have focused our attention to $|g|$ small enough such that
for attractive condensates the ground state is always coupled
through the stirring strength and velocity to all other states.
This is not the case at $g\ll0$ (see Appendix~\ref{app:spectrum}), where new types of solutions appear.
In this case, the ground state does not have a flat density.
It can be stirred with increasing velocity, analogously to the 
above excitation processes,
 but the final states are also not flat.

Note that in this section we considered the adiabatic following of particular eigenstates of the system. That is, we assume that the adiabaticity condition is fulfilled and no other eigenstates are populated during the stirring \cite{Messiah_1976}. In particular, if we consider only the closest energy eigenstates of the system, the adiabatic condition will depend on the gap opened by the barrier. This is particularly important around the swallowtails where the gap is small. In a simplified two-state model the adiabaticity condition can be calculated analytically and is given by $\frac{1}{2}\left|\frac{d\lambda}{dt}\Delta E - \lambda\frac{d\Delta E}{dt}\right|\ll \left( \lambda^2 + \Delta E ^2 \right)^3/2$~\cite{Eberly_1987,Bergmann_1987}, with $\Delta E$ being the energy difference between eigenstates and $\lambda$ the tunneling amplitude between the two states.

\subsection{Hysteresis cycles}
\label{sec:hysteresis}

Hysteresis due to a rotating weak link in a Bose gas was
first experimentally observed in~\cite{eckel14}. These hysteresis cycles are understood
in terms of (downward) swallowtail diagrams~\cite{mueller02}, and
their widths were numerically computed in~\cite{munoz15}.
Here we discuss how for a delta type link the widths
and heights of the hysteresis cycles,
$\Delta \Omega$ and $\Delta J$,
can be computed using the exact spectrum presented in this work.
Our simplified 1D GPE model thus yields a minimal expression of the
hysteresis results found in~\cite{eckel14}, providing  insight
into the fundamental hysteresis mechanism from a more analytical point of view.
In this experimental setup, a blue-detuned laser beam creates an effective repulsive potential,
which is rotated or stopped around the ring trap to create or destroy currents.
We follow an analogous protocol through cycles which contain adiabatic
excitation paths and spontaneous decay. The adiabaticity condition, discussed
previously, is related to the energy gap between the present state and the closest energy eigenstate,
which in turn depend on the weak link strength. This condition does not represent
a major limiting factor in these experiments, in which the barrier strength can
be tuned to increase or decrease the energy gaps.
Differences between 1D GPE results and experiments, typically performed in 3D
setups where transverse excitations can play a substantial role, might still appear.
Our model, however, is capable of describing some of the main qualitative features
found in experiments.

Fig.~\ref{fig:st3x3} proves useful to
illustrate the main features of hysteresis.
On the one hand, it shows that only repulsive condensates present
downward swallowtail structures, and therefore the associated 
hysteresis cycles only exist for $g>0$.
This is because when the critical velocity is reached
at the tip of an upward swallowtail, the state with lower
energy to which the condensate decays belongs to a lower set
of concatenated swallowtails. This effectively impedes to 
excite the condensate back to the upper swallowtail 
through any adiabatic variation of $\Omega$, and therefore
to close the cycle.
On the other hand, hysteresis cycles on repulsive condensates
are not characteristic
of a stirring of the ground state, where the condensate
undergoes a transition $\Delta J\simeq 1$.
Stirring of excited states also present hysteresis,
each excited state implying a different width $\Delta \Omega$
and height $\Delta J$,
features not discussed in previous works.
Moreover, these cycles can be analyzed in terms
of the nonlinearity and link's strength.
In general, the range of velocities limited by $\Omega_{c}$,
and the associated adiabatic widths of the swallowtails and hysteresis cycles, $\Delta \Omega$, 
become smaller as larger transitions $\Delta J$ are considered,
as $g$ decreases, or as the link's magnitude $\alpha$ becomes stronger.
This qualitatively agrees with the experimental results found 
in Fig. 3 of~\cite{eckel14}, where the bottom and top widths of the cycles,
in which the current varies slightly, and the critical velocities
decrease as stronger barriers are considered.
On the other hand, the precise dependence of the current on the velocity
in the non-adiabatic transitions, in which the current is observed to
increase more abruptly, is beyond the scope of our model.
In the limit $\alpha\to0$, the heights and widths of the cycles
become $\Delta J=n$, $\Delta \Omega=2\,\tilde{\Omega}_n-n$,
with integer $n$.

\begin{figure}[t]
\centering
\includegraphics[width=.48\textwidth,height=.274\textwidth]{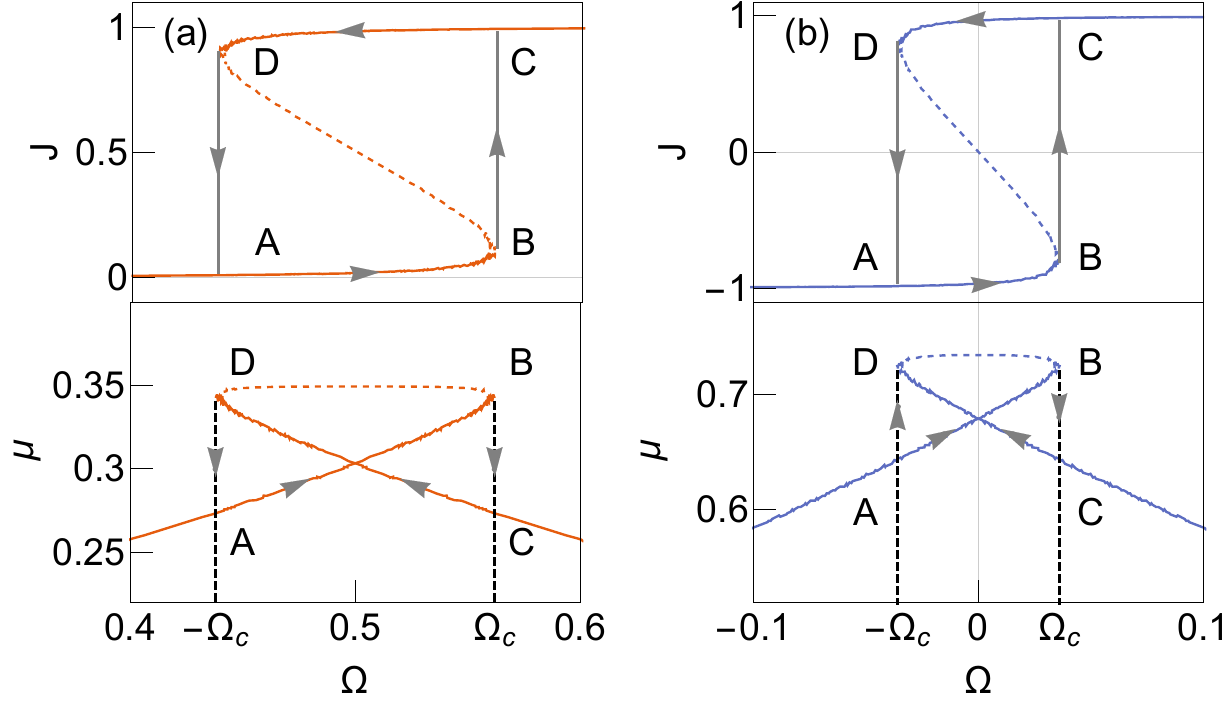}
\caption{
Hysteresis cycles in terms of currents and stirring velocities 
(top plots) and corresponding energy diagrams in form of swallowtails
(bottom plots) for $\alpha=\frac18$.
(a) Stirring of the ground state up to $\Omega_{c}$, $A\to B$,
transition to $J\simeq 1$ through the gray line (vertical arrow), $B\to C$,
and stirring in opposite direction, $C\to D$, where the condensate
decays to the original state, $D\to A$.
(b) When a link is set in a vortex state, two solitons are produced,
but the hysteresis paths are analogous to the ground state case. 
In this cycle, the paths are cut short at much smaller critical velocities.
$\mu$, $J$, and $\Omega$ are in natural units.
}
\label{fig:hysteresis}
\end{figure}

In Fig.~\ref{fig:hysteresis} we present two hysteresis cycles,
one corresponding to stirring the ground state, where $\Delta J\simeq1$,
 and another associated to setting a link in a vortex state,
and then stirring, effectively coupling $J\simeq-1$ and $J\simeq1$ vortex states.
Both cycles involve adiabatic paths in which
the condensate is stirred up to the corresponding critical velocity.
Passed these velocities, the condensate is assumed to decay to the next vortex state,
that is, to the lower branch of the same swallowtail structure,
increasing the current in roughly one and two units of angular momentum, respectively.
Note that these paths, shown in gray (vertical arrows) in Fig.~\ref{fig:hysteresis}, are plotted straight
and vertically, just to assume the simplest case.
Then the condensate is stirred in the opposite direction,
 where the same process is repeated,
returning thus to the original state.
The critical velocities limiting the stirring of the ground state are found to coincide
with $\sqrt{g \rho_0}$, where $\rho_0$ is the density at the lowest point,
as in~\cite{munoz15}.
For excited states, however, 
these velocities slightly depart from the sound velocity at the low density
region.

In~\cite{perezobiol20}, an extensive discussion on the stability of 
swallowtail diagrams was performed, including a precise formula
to compute the critical velocities. Solutions constituting the upper parts
of swallowtails were found unstable with respect to Bogoliubov perturbations,
 while the lower parts of the ground and
first excited state were fully stable for repulsive links.
This implies that the initial paths of these hysteresis cycles, corresponding
to the lower part of the swallowtails, constitute
a set of stable solutions. Once the tip of the swallowtail is reached
through the lower part, and a critical velocity is encountered,
the set of stable solutions merges with the set of unstable ones.
At this point, the condensate undergoes a non-adiabatic transition towards
the other lower branch of the swallowtail.
The precise dynamics of these instabilities is beyond the capabilities 
of the current model. However, knowing the precise values of the critical velocities,
 and the corresponding solutions in terms of elliptic functions,
might allow for an easier identification of the physical mechanisms
behind the bifurcation instabilities, with either finite temperature, time-dependent
computations~\cite{karpiuk12,katsimiga18}, or through more general instability criteria
such as the enhancement of dynamical density fluctuations~\cite{kato10,watabe13}.

\subsection{Auxiliary potential and phase imprinting}
\label{sec:auxiliary}

\begin{figure*}[t]
\centering
\includegraphics[width=.96\textwidth,height=0.342\textwidth]{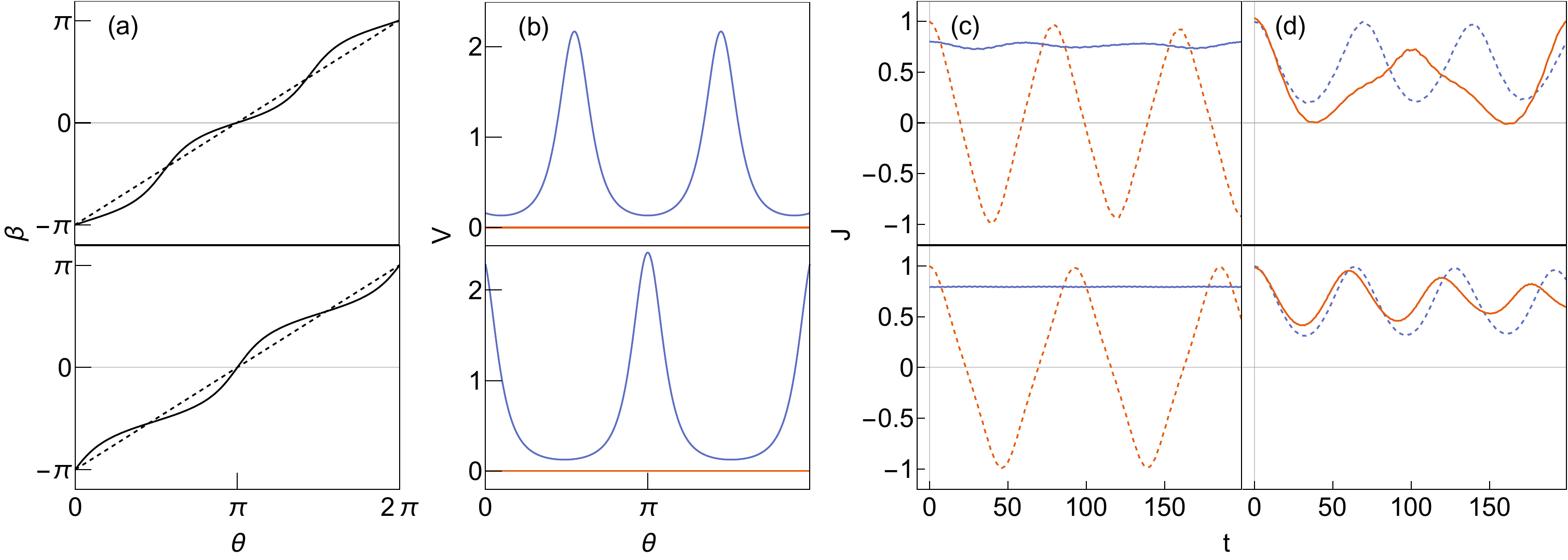}
\caption{Top plots correspond to $g=-1$ and bottom ones to $g=1$.
(a): linear ($2\pi x$, dashed lines) and nonlinear ($\beta_s(\theta)$,
solid lines) phases.
(b): auxiliary potential $V(\theta)=V_{aux}(\theta)$ (blue/upper line) and $V(\theta)=0$ (red/lower line).
(c): current evolution after regular phase imprinting (dashed red),
and after applying the auxiliary potential and imprinting
the nonlinear slope (solid blue). The protocol is carried out on the ground state 
of the GPE with a link of strength $\alpha=0.3$.
(d) intermediate cases in which only  the auxiliary potential (dashed blue) or
the nonlinear phase imprinting (solid red) are used.
Units are dimensionless with $R=M=\hbar=1$.
}
\label{fig:vaux}
\end{figure*}

The adiabatic paths presented so far involve
the movement of the link or a stirring.
An alternative procedure to excite the condensate
is to imprint a phase through an electromagnetic field \cite{perrin18}.
Phase imprinting provides a fast way to excite the condensate, but when the ring contains
a static defect, which might happen naturally in experiment,
the state obtained is not stationary, and one encounters
BJJ oscillations \cite{polo19}, or other nonlinear effects if $g$ is large enough.
These oscillations can be understood intuitively through simplified hydrodynamics considerations.
If a linear phase is imprinted, all the atoms throughout
the annular trap acquire the same momentum, which implies
a smaller current at the low density region
created by the defect.
The condensate thus accumulates at the side of the defect,
slows down, and bounces to the other side of the barrier.
This effect can be partially reduced by
imprinting a nonlinear phase such that a larger kick is provided
to the condensate at the low density region,
producing thus a current which is roughly stationary, i.e.,
$J(\theta)=\rho(\theta)\beta'(\theta)\simeq{\rm constant}$.
This idea can be quantitatively analyzed since the excited stationary states
with current $J\simeq n$ and a delta defect are well known in terms of Jacobi elliptic functions.
One can in principle imprint the phase of these states
on the ground state, but the unperturbed density
would still differ from the densities of stationary ones,
which, apart from the dip produced by the link,
contain other gray solitons.
Therefore, the final states would not be stable.
To solve this, and taking advantage of adiabatic processes,
we design a protocol to
produce the density of the desired excited state 
through an auxiliary potential \cite{Schnelle,Henderson20}, and leaving the link fixed.
Once this condensate's density
is obtained, the phase is imprinted and the auxiliary potential is turned off.

More explicitly, if the final stationary solution we want to obtain, in presence of the delta potential,
is $\psi_s(\theta)=r_s(\theta)e^{i \beta_s(\theta)}$, we first set an auxiliary potential $V_{aux}$ such
that the ground state is $\psi_g=r_s(\theta)$, and therefore satisfies
\begin{align}
 \tilde{\mu} r_s=&
-\frac12 r_s''
+g r_s^3+V_{aux}\,r_s.
\end{align}
On the other hand, the final excited state we want to build is determined by
\begin{align}
 \mu r_s=&
-\frac12 [r_s''-r_s \beta_s'^2+i(r_s\beta_s''+2r_s'\beta_s')]
+g r_s^3.
\end{align}
The imaginary part is zero as long as $\beta_s'=\frac{\gamma}{r_s^2}$,
with $\gamma$ being a constant representing the current.
Subtracting both equations, and neglecting the constant $\tilde{\mu}-\mu$, we find
\begin{align}
 V_{aux}(\theta)=\frac12 \beta_s'(\theta)^2=
\frac{\gamma^2}{2\,r_s(\theta)^4}.
\end{align}
The protocol then consists in gradually turning on $V_{aux}(\theta)$, for example by increasing 
its overall factor from zero to one, and then turn if off while phase imprinting $\beta_s(\theta)$.
The stationary state $r_s(\theta)e^{i\beta_s(\theta)}$ with current $J\simeq n >0$
is thus accessed.

This protocol is reproduced through simulations in the time-dependent
GPE, as shown in Fig.~\ref{fig:vaux}, obtaining a steady current $J\simeq1$.
This is in contrast with regular phase imprinting, which for
large enough defects, in our tested case $\alpha=0.3$, produces BJJ oscillations,
as also shown in the figure.
Apart from the protocol described above, we study
the evolution of the current when only the auxiliary potential
is used ---only the density of the stationary current with a delta is imitated,
and a regular slope $2\pi x$ is imprinted---,
and when only the non-linear phase imprinting is used on the ground state
with a delta. 
We observe that in all cases, both the auxiliary potential
and the nonlinear phase imprinting, serve independently
to produce more self-trapping in the final state.

\section{Conclusions}
\label{sec:conclusions}

We have described the spectrum of solitons moving at constant velocity
in a ring condensate, either freely, or being dragged
by a weak link, and with either attractive or repulsive interactions.
Their energies, densities, and currents have been
thoroughly analyzed in terms of the link's strength $\alpha$ and velocity
$\Omega$, and found that all states are coupled at $\alpha=0$.
At $\alpha\neq0$, the steady dragged solitonic solutions exist only in certain ranges of link velocities,
periodic in $\Omega$, and the midpoints of which, at integer and half-integer
velocities, correspond to dark
solitonic states.

By studying how the different stationary states are connected
through an adiabatic variation of $\alpha$ and $\Omega$, we have laid out three different
methods to modify the state of the condensate in a controlled manner.
The first, involves a purely adiabatic variation of the link's strength and velocity.
Setting a link and then stirring by increasing its velocity,
allows to excite the ground state of attractive condensates, but
in all other cases critical velocities are encountered, and current variations
are limited to $|\Delta J|\lesssim 1$. To access excited states,
the link must be set while rotating at a finite velocity.
Secondly, we have considered processes in which the link moves but
its strength is kept fixed. In this case, the link's velocity surpasses
the critical one and the condensate is assumed to decay to the immediate lower state.
For repulsive condensates, these paths, consisting in both an adiabatic
excitation part and a non-adiabatic decay, can be closed by moving the
weak link in both directions, and effectively producing hysteresis cycles.
Here, we have shown that these hysteresis cycles can also be produced in excited states,
although they are limited by different critical velocities,
and that hysteresis cycles cannot exist for attractive condensates.
Finally, we have made use of an auxiliary potential to adiabatically
modify the ground state density, and to then imprint a nonlinear phase
while the potential is turned off. The auxiliary potential and phase are
precisely designed such that the state produced is an excited but stationary
state, and no BJJ oscillations are found.

This work illustrates, from an analytical point of view, the physical mechanisms involved
in the production of currents in weakly interacting Bose gases in a ring trap.
It also provides a theoretical description which allows for further exploration of the system,
including ground states as well as excited states.

\begin{acknowledgments}
A.P-O. and T.C. acknowledge the support by KUT presidential grant
at Research Institute, Kochi University of Technology, Japan.
J.P. acknowledges Okinawa Institute of Science and Technology Graduate University
and also the JSPS KAKENHI Grant Number 20K14417.
\end{acknowledgments}

\appendix
\section{Spectrum}
\label{app:spectrum}

The spectrum of normalized solutions is more easily analyzed in the delta
comoving frame, which is determined by
Eqs.~(\ref{eq:gpf})-(\ref{eq:bc2}).
The solutions of these equations
are parametrized through  a density $\rho$ and a phase $\beta$,
$\phi(\theta)=\sqrt{\rho(\theta)}e^{i\beta(\theta)}$,
and are given in analytical form in terms
of Jacobi elliptic functions ($F$), $\rho(\theta)=A+B\,F^2(k(\theta-\theta_0),m)$,
and where $\rho(\theta)\beta'(\theta)=\gamma$ is  constant.
In our case, we set the Jacobi function $F=dn$.
$A$, $B$ and $\theta_0$ depend on the frequency $k\geq0$ and elliptic modulus $m\in[0,1]$.
$\rho(\theta)$ in general oscillates around a finite value, has no zeros,
and is smooth except for the derivative jump due to the delta
at $\theta=0$.
In the limits $A\to0$ and $B\to0$, these solutions become dark solitons
and plane waves, respectively.
 In  the following we illustrate the main
features of the spectrum according to Fig.~\ref{fig:st3x3}.

In the linear and free case, $g=0$ and $\alpha=0$, the solutions are plane waves
or vortex states, with the chemical potential quantized by periodic
conditions and given by $\mu=\frac{n^2}{2}$.
Each parabola in Fig.~\ref{fig:st3x3} (d) represents the energy
of each of these states,
$\mu=\frac12(\Omega+n)^2$,
as measured by an observer moving around the ring at constant velocity
 $\Omega$. In the lab frame each parabola represents
the same solution with energy
$\frac{n^2}{2}$.

When atomic interactions are finite, $g\neq0$, and no link is present, $\alpha=0$,
the condensate is governed by the GPE with periodic boundary conditions,
 which also has as solutions
plane waves. Their energy includes the same kinetic term
as in the linear case, but also a potential term
which shifts the energy parabolas upward and downward
for the repulsive and attractive cases,
$\mu=\frac{g}{2\pi}+\frac12(\Omega+n)^2$.
Moreover, there are new sets of solutions, consisting in gray
solitons moving at constant velocity $\Omega$.  
Their energies as a function of $\Omega$, in the frame of reference
of the moving solitons, are shown in Fig.~\ref{fig:st3x3} (a) and (g)
as the curves crossing between the plane wave parabolas.
The middle points of these crossing lines, marked as black dots,
correspond to dark solitons. Even number of dark solitons move
at velocities $\Omega=n$, while odd number of dark solitons
 travel at $\Omega=n+\frac{1}{2}$, with $n$ an integer.
These waves are always non-moving with respect to the condensate.
As the gray solitons become shallower, their
velocities depart from $\Omega=n,n+\frac12$, until
the densities become completely flat and these solutions
merge with the plane waves.
These sets of solutions can be understood as energy
bands in the lab frame. Each band consists in
solutions with a fixed number of solitons with velocities
ranging from $m$ to $\Omega=m\pm n$.
A more rigorous analysis of the first
band of gray solitonic solutions can be found in~\cite{sacchetti20}.

The energies of the gray solitons increase and decrease
with $g$, and this implies the appearance of new static solutions
($\Omega=0$) as $|g|$ grows larger \cite{carr002}, in particular
of new ground states for $g<0$.
In this article we have focused on $|g|$ small enough so that the ground
state for attractive condensates stays coupled to the rest of the spectrum.
To illustrate how the spectrum qualitatively depends on $g$,
we consider two particular cases, one for attractive condensates
and one for repulsive ones.
First, as $g$ decreases from $g=0$, the lowest of these crossing lines
(blue lines in Fig.~\ref{fig:st3x3} (g)) move downward, until their left
and right limits coincide with the bottom points of the parabolas,
at $\tilde{\Omega}_1=0$, $g=-\frac{\pi}{2}$.
For $g<-\frac{\pi}{2}$, the ground state as a function of $\Omega$ forms
a continuous line uncoupled from the rest of parabolas.
In general, new uncoupled states appear at $g<-\frac{n^2\pi}{2}$,
with integer $n>0$.
Another example is the appearance of a new second excited state
as $g$ grows and the red solitonic line in Fig.~\ref{fig:st3x3} (a)
crosses the axis $\Omega=0$. More precisely, at $g=\frac{3\pi}{2}$,
 such that the left limit coincides with the vertical axis, $1-\tilde{\Omega}_1=0$,
a new solution appears at $\Omega=0$ between the first vortex state
and the first dark solitonic solution.

\begin{figure}[t]
\centering
\includegraphics[width=.48\textwidth,height=.335\textwidth]{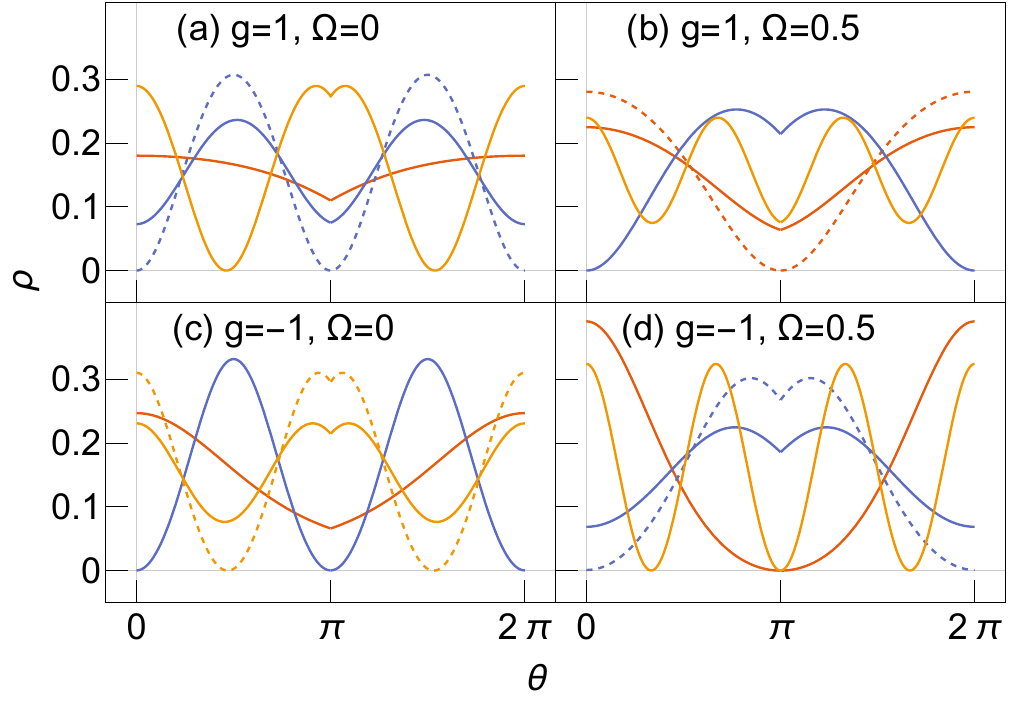}
\caption{Densities for the ground state and first excited states given 
a link of strength $\alpha=\frac14$ and velocities and nonlinearities $g=\pm 1$,
$\Omega=0,0.5$. Colors (shades of grey) match the ones of their
respective energies in Fig.~\ref{fig:st3x3}, and dashed lines correspond
to upper parts of downward swallowtails, or bottom parts
of upward ones. In this figure, the delta
is placed at $\theta=\pi$ for visualization purposes. In the rest
of the article the delta conditions are at $\theta=0,2\pi$.
All units are dimensionless and $R=M=\hbar=1$.
}
\label{fig:densities}
\end{figure}

As a link is turned on, $\alpha>0$, the spectrum of plots (a), (d), and (g)
described above splits into a set of diagrams
separated by a gap. For finite $g$, these diagrams have
the shape of downward ($g>0$) and upward ($g<0$) swallowtails.
The gap among the swallowtails grows for larger
delta strengths and nonlinearities.
Each set of concatenated swallowtail diagrams represents
a set of solutions with a fixed number of solitons.
The densities of the lowest set of swallowtails have only
the downward kink created by the delta.
Then, each superior set has, apart from the dip in the density
produced by the link, one more gray soliton.
The middle points of these diagrams still represent dark solitons,
each previous dark solitonic solution at $g=0$ now split into two.
In the solution with higher energy, the dark soliton coinciding
with the delta corresponds to a derivative jump,
satisfying delta conditions.
The solution with lower energy consists in a periodic and smooth
wave such that one of the zeros coincides with the position of the
link. Solutions of this type trivially satisfy delta conditions
for any $\alpha$, since the derivatives and the function
at the position of the delta are zero. In Fig.~\ref{fig:st3x3},
this means the black dots in the red lines, the ones in the blue
lines at $\Omega=n$, and the ones at the orange lines at $\Omega=n+\frac{1}{2}$,
have the same $\mu$ across the panels in each row.
As a sample, the densities of these pairs of dark solitonic trains,
and the ground and other excited states,
are plotted in Fig.~\ref{fig:densities} for 
$\alpha=\frac14$, $g=-1,1$ and $\Omega=0,0.5$.

We herewith have thoroughly
described the set of solitonic solutions in correspondence
to Fig.~\ref{fig:st3x3}.
To sum up, the solutions for $\alpha=0$ consist of vortex states with
current $J=n$,  of $m$ dark solitons moving at
$\Omega=\frac{m}{2}+ n$, and of $m$ gray solitons
traveling at $\Omega\in(\frac{m}{2}+ n-|\tilde{\Omega}_m-\frac{m}{2}|,\frac{m}{2}
+ n+|\tilde{\Omega}_m-\frac{m}{2}|)$, with integers $n$, $m$.
For a rotating link, the dragged solutions comprise
trains with $m+1$ gray solitons coupling solutions with 
$m$ and $m+1$ dark solitons moving at 
$\Omega=\frac{m}{2}+n$ and $\Omega=\frac{m}{2}+n+\frac12$, respectively,
and limited by critical velocities $\Omega_c$.


\begin{thebibliography}{9}

\bibitem{Amico_2017}
L. Amico, G. Birkl, M. Boshier and L.-C. Kwek
New J. Phys. {\bf 19}, 020201 (2017).

\bibitem{ryu07}
C. Ryu, M. F. Andersen, P. Clad\'e, V. Natarajan, K. Helmerson, and W. D. Phillips,
Phys. Rev. Lett. {\bf 99}, 260401 (2007).



\bibitem{moulder12}
S. Moulder, S. Beattie, R. P. Smith, N. Tammuz, and Z. Hadzibabic,
Phys. Rev. A {\bf86}, 013629 (2012).

\bibitem{beattie13}
S. Beattie, S. Moulder, R. J. Fletcher, and Z. Hadzibabic,
Phys. Rev. Lett. {\bf110}, 025301 (2013).

\bibitem{wright13pra}
K. C. Wright, R. B. Blakestad, C. J. Lobb, W. D. Phillips, and G. K. Campbell
Phys. Rev. A {\bf88}, 063633 (2013).

\bibitem{eckel14}
S. Eckel, J. G. Lee, F. Jendrzejewski, N. Murray, C. W. Clark, C. J. Lobb,
W. D. Phillips, M. Edwards, and G. K. Campbell,
Nature (London) {\bf506}, 200 (2014).


\bibitem{ryu13}
C. Ryu, P. W. Blackburn, A. A. Blinova, and M. G. Boshier,
Phys. Rev. Lett. 111, 205301 (2013).


\bibitem{hallwood10}
D. W. Hallwood, T. Ernst, and J. Brand, Phys. Rev. A {\bf 82}, 063623 (2010);
D. Solenov and D. Mozyrsky, Phys. Rev. A {\bf 82}, 061601(R) (2010);
A. Nunnenkamp, A. M. Rey, and K. Burnett, Phys. Rev. A {\bf 84}, 053604 (2011);
D. Solenov and D. Mozyrsky, J. Comput. Theor. Nanosci. {\bf 8}, 481 (2011).

\bibitem{schenke11}
C. Schenke, A. Minguzzi, and F. W. J. Hekking, Phys. Rev. A 84, 053636 (2011).

\bibitem{amico14}
L. Amico, D. Aghamalyan, F. Auksztol,H. Crepaz, R. Dumke, and L.-C. Kwek, Sci. Rep. 4, 4298 (2014).


\bibitem{Hou_2017}
Junpeng Hou, Xi-Wang Luo, Kuei Sun, and Chuanwei Zhang
Phys. Rev. A {\bf 96}, 011603(R) (2017).

\bibitem{perrin20}
Y. Guo, R. Dubessy, M. G. de Herve, A. Kumar, T. Badr, A. Perrin, L. Longchambon, and H. Perrin,
Phys. Rev. Lett. {\bf 124}, 025301 (2020).



\bibitem{Wright_2013}
K. C. Wright, R. B. Blakestad, C. J. Lobb, W. D. Phillips, and G. K. Campbell
Phys. Rev. Lett. {\bf 110}, 025302 (2013).


\bibitem{Amico_2015}
L. Amico, D. Aghamalyan, F. Auksztol, H. Crepaz, R. Dumke and L.-C. Kwek 
Scientific Reports {\bf 4}, 4298 (2015).


\bibitem{Aghamalyan_2015}
D. Aghamalyan, M. Cominotti, M. Rizzi, D. Rossini, F. Hekking, A. Minguzzi, L.-C. Kwek and L. Amico,
New J. Phys. {\bf 17}, 045023 (2015).


\bibitem{piazza09}
F. Piazza, L. A. Collins, and A. Smerzi,
Phys. Rev. A {\bf 80}, 021601(R) (2009).


\bibitem{ramanathan11}
A. Ramanathan, K. C. Wright, S. R. Muniz, M. Zelan, W. T. Hill, C. J. Lobb, K. Helmerson, W. D. Phillips, and G. K. Campbell,
Phys. Rev. Lett. {\bf 106}, 130401 (2011).

 \bibitem{piazza13}
F. Piazza, L. A. Collins, and A. Smerzi,
J. Phys. B: At. Mol. Opt. Phys. {\bf 46}, 095302 (2013).

\bibitem{wright13prl}
K. C. Wright, R. B. Blakestad, C. J. Lobb, W. D. Phillips, and G. K. Campbell,
Phys. Rev. Lett. {\bf 110}, 025302 (2013).





\bibitem{dalibard11}
J. Dalibard, F. Gerbier, G. Juzeli\=unas, and P. \"Ohberg,
Rev. Mod. Phys. {\bf 83}, 1523 (2011).

\bibitem{perrin18}
A. Kumar, R. Dubessy, T. Badr, C. De Rossi, M. de Go\"er de Herve, L. Longchambon, and H. Perrin
Phys. Rev. A {\bf97}, 043615 (2018).



\bibitem{polo19}
J. Polo, R. Dubessy, P. Pedri, H. Perrin, and A. Minguzzi
Phys. Rev. Lett. {\bf 123}, 195301 (2019).


\bibitem{kanamoto09}
R. Kanamoto, L. D. Carr, and M. Ueda
Phys. Rev. A {\bf 79}, 063616 (2009).

\bibitem{fialko12}
O. Fialko, M.-C. Delattre, J. Brand, and A. R. Kolovsky,
Phys. Rev. Lett. {\bf108}, 250402 (2012).

\bibitem{li12}
Y. Li, W. Pang, and B. A. Malomed,
Phys. Rev. A {\bf 86}, 023832 (2012).

\bibitem{munoz19}
A. Mu\~noz Mateo, V. Delgado, M. Guilleumas, R. Mayol, and J. Brand,
Phys. Rev. A {\bf 99}, 023630 (2019).





\bibitem{baharian13}
S. Baharian and G. Baym,
Phys. Rev. A {\bf 87}, 013619 (2013).

\bibitem{munoz15} 
A. Mu\~noz Mateo, A. Gallem\'i, M. Guilleumas, and R. Mayol,
Phys. Rev. A {\bf 91}, 063625 (2015).

\bibitem{kunimi18} 
M. Kunimi and Y. Kato,
Phys. Rev. A {\bf 91}, 053608 (2015).




\bibitem{carr002} 
L. D. Carr, C. W. Clark, and W. P. Reinhardt,
Phys. Rev. A 62, 063610 (2000);
Phys. Rev. A {\bf 62}, 063611 (2000).

\bibitem{seaman05} 
B. T. Seaman, L. D. Carr, and M. J. Holland,
Phys. Rev. A {\bf 71}, 033609 (2005).

\bibitem{hakim97}
V. Hakim,
Phys. Rev. E {\bf 55}, 2835 (1997).

\bibitem{pavloff02}
N. Pavloff,
Phys. Rev. A {\bf 66}, 013610 (2002).

\bibitem{cominotti14}
M. Cominotti, D. Rossini, M. Rizzi, F. Hekking, and A. Minguzzi,
Phys. Rev. Lett. {\bf 113}, 025301 (2014).

\bibitem{shamriz18}
E. Shamriz and B. A. Malomed
Phys. Rev. E {\bf 98}, 052203 (2018).


\bibitem{perezobiol19}
A. P\'erez-Obiol and T. Cheon,
J. Phys. Soc. Jpn. {\bf 88}, 034005 (2019).

\bibitem{perezobiol20}
A. P\'erez-Obiol and T. Cheon
Phys. Rev. E {\bf 101}, 022212 (2020).



\bibitem{kunimi19}
M. Kunimi and I. Danshita,
Phys. Rev. A {\bf 100}, 063617 (2019).


\bibitem{takahashi16}
D. A. Takahashi,
Phys. Rev. E {\bf 93}, 062224 (2016).

\bibitem{feng19}
B.-F. Feng, L. Ling, and D. A. Takahashi
Stud. App. Math. {\bf 144} 46-101 (2019).


\bibitem{mueller02}
E. J. Mueller,
Phys. Rev. A {\bf 66}, 063603 (2002).

\bibitem{Messiah_1976}
A. Messiah. Quantum Mechanics, 2, North-Holland, Amsterdam (1976).

\bibitem{Eberly_1987}
L. Allen and J. H. Eberly. Optical resonance and two-level atoms, Dover Publications, New York, (1987).

\bibitem{Bergmann_1987}
N. V. Vitanov, T. Halfmann, B. W. Shore, and K. Bergmann, Annual Review of Physical Chemistry 52, 763 (2001).



\bibitem{karpiuk12}
T. Karpiuk, P. Deuar, P. Bienias, E. Witkowska, K. Pawlowski, M. Gajda, K. Rzazewski, M. Brewczyk,
Phys. Rev. Lett. {\bf 109}, 205302 (2012).

\bibitem{katsimiga18}
G. C. Katsimiga, S. I. Mistakidis, G. M. Koutentakis, P. G. Kevrekidis, and P. Schmelcher,
Phys. Rev. A {\bf 98}, 013632 (2018).

\bibitem{kato10}
Y. Kato and S. Watabe,
Phys. Rev. Lett. {\bf 105}, 035302 (2010).


\bibitem{watabe13}
S. Watabe and Y. Kato,
Phys. Rev. A {\bf 88}, 063612 (2013).



\bibitem{Schnelle}
S. K. Schnelle, E. D. van Ooijen, M. J. Davis, N. R. Heckenberg, and H. Rubinsztein-Dunlop,
Opt. Express, {\bf 16}, 1405, (2008).

\bibitem{Henderson20}
K. Henderson, C. Ryu, C. MacCormick, and M. G. Boshier,
New J. Phys., {\bf 11}, 043030, (2009).


\bibitem{sacchetti20}
A. Sacchetti, J. Phys. A {\bf 53}, 385204 (2020).




\end{thebibliography}
\end{document}